# Radiation Resilience of β-Ga₂O₃ Schottky Barrier Diodes Under High Dose Gamma Radiation

Saleh Ahmed Khan[1, a)], Sudipto Saha[2, a)], Uttam Singisetti[2], A F M Anhar Uddin Bhuiyan[1, b)]

[1]Department of Electrical and Computer Engineering, University of Massachusetts Lowell, MA, USA

[2] Department of Electrical Engineering, University at Buffalo, Buffalo, NY, USA

[a)] *S. A. Khan and S. Saha contributed equally to this work*

[b)] Corresponding Author Email: anhar_bhuiyan@uml.edu

## Abstract

A systematic investigation of the electrical characteristics of β-Ga₂O₃ Schottky barrier diodes (SBDs) has been conducted under high-dose $^{60}$Co gamma radiation, with total cumulative doses reaching up to 5 Mrad (Si). Initial exposure of the diodes to 1 Mrad resulted in a significant decrease in on-current and an increase in on-resistance compared to the pre-radiation condition, likely due to the generation of radiation-induced deep-level acceptor traps. However, upon exposure to higher gamma radiation doses of 3 and 5 Mrad, partial recovery of the device performance occurred, attributed to a radiation annealing effect. Capacitance-voltage (C-V) measurements showed a decrease in net carrier concentration in the β-Ga₂O₃ drift layer, from ~3.20 × 10¹⁶ to ~3.05 × 10¹⁶ cm⁻³ after 5 Mrad irradiation. Temperature-dependent I-V characteristics showed that 5 Mrad irradiation leads to a reduction in both forward and reverse current across all investigated temperatures ranging from 25 to 250°C, accompanied by slight increases in on-resistance, ideality factors, and Schottky barrier heights. Additionally, a slight increase in reverse breakdown voltage was observed post-radiation. Overall, β-Ga₂O₃ SBDs exhibits high resilience to gamma irradiation, with performance degradation mitigated by radiation-induced self-recovery, highlighting its potential for radiation-hardened electronic applications in extreme environment.

## I. Introduction

Radiation-hardened electronic devices are essential for space missions, nuclear power plants, and particle accelerators, requiring robust engineering to resist ionizing radiation while maintaining optimal performance. Wide and ultrawide bandgap materials show superior radiation tolerance compared to Si-based devices due to their larger bandgaps. Among these, β-Ga2O3 is a promising candidate for radiation-resistant electronics due to its ultrawide bandgap energy (~4.8 eV) and stable crystal structure [1-3]. The strong gallium-oxygen bonding in β-Ga2O3 results in higher displacement energies, enhancing the lattice's robustness against external perturbations, allowing devices to endure substantial radiation exposure without performance degradation. Additionally, β-Ga2O3 features a high breakdown field strength of 8 MV/cm, large Baliga's figure of merit, controllable n-type doping, and availability of high-quality large native substrates, making it a scalable and cost-effective alternative to other wide bandgap materials such as GaN or SiC for power electronics in harsh environments [4-17].

Understanding the radiation tolerance of semiconductor materials is crucial for developing electronic devices that can endure ionizing radiation in extreme environments. This is because radiation can cause dislocations or defects in the semiconductor lattice, degrade dielectric layers, and impair metal contacts, thereby compromising performance. The extent of this damage also varies with the type and intensity of the radiation: heavy particles such as neutrons or heavy ions tend to create permanent disordered regions within the semiconductor lattice, whereas gamma rays passing through the material can create traps or charged regions without significantly altering the lattice structure [17-19]. Previous studies have extensively investigated the radiation tolerance of β-Ga$_2$O$_3$ devices under various radiation environments. For instance, studies on the impact of penetrating neutron radiation on β-Ga$_2$O$_3$ SBDs have revealed the formation of bulk traps within the drift layer, leading to changes in device performance [20-22]. Similarly, proton radiation studies have demonstrated increased carrier compensation due to proton-induced damage [23-25]. Additionally, the effects of heavy ion irradiation on β-Ga$_2$O$_3$ Schottky diodes have also been investigated [26], revealing an inherently low susceptibility to radiation-induced material degradation. However, the introduction of acceptor defects at the interface and within the drift layer was found to trap carriers, thereby reducing the net carrier concentration in the epitaxial layer. The influence of gamma radiation on semiconductors is also of significant interest because their interaction mechanism differs from that of other charged particles. Energy loss mechanisms for gamma rays in semiconductor materials can be categorized into ionizing and non-ionizing types, with the latter being lower for gamma rays than for heavy charged particles or ions. Gamma ray interactions with semiconductors vary by energy level: the photoelectric effect dominates at <1 MeV, e-h pair production at >10 MeV, and the Compton effect, creating secondary electrons, at 1-10 MeV [27]. These secondary electrons can influence trapped and interface charges, altering device performance, and can create Frenkel pairs, displacing lattice atoms. The effects of $^{60}$Co gamma irradiation on GaN diodes have been studied previously at doses up to 21 Mrad (Si) [28]. The results showed a noticeable increase in the Schottky Barrier Height (SBH) and significant degradation in reverse I-V characteristics, while the impact on forward I-V characteristics was minimal. These changes were primarily attributed to dislocations in the GaN epilayer and radiation-induced defects at the metal-semiconductor interface. While GaN shows relatively low susceptibility to radiation-induced material degradation, its radiation hardness is often limited by the vulnerability of the metal-GaN interface and the high density of dislocations, which negatively affect reverse-bias characteristics. Another study on total ionizing dose exposure showed that GaN quasi-vertical Schottky diodes exhibited resilience to gamma radiation up to 1 Mrad [29]. A report on 4H-SiC Schottky barrier diodes indicated that these devices tolerate radiation doses up to 1 Mrad, but higher doses lead to an increase in negative oxide charge accumulation. [30]. While extensive studies have investigated gamma irradiation damage in GaN, SiC and GaAs devices [28-37], research on the influence of gamma radiation on the structural and electrical properties of $\beta$-$Ga_2O_3$ materials and devices remains limited. Potential lattice defects in $\beta$-$Ga_2O_3$ include deep acceptor Ga- and shallow donor O- vacancies, with Ga vacancies forming more readily during radiation exposure [38, 39]. For instance, an investigation into 1.6 MGy ($SiO_2$) gamma radiation on $\beta$-$Ga_2O_3$ MOSFETs demonstrated that, although $\beta$-$Ga_2O_3$ exhibits considerable intrinsic radiation tolerance, the overall radiation hardness of these devices was limited by radiation-induced gate leakage and drain current dispersion, which were linked to dielectric damage and interface charge trapping [40]. Additionally, exposure of $\beta$-$Ga_2O_3$ rectifiers to 100 krad (Si) gamma radiation resulted in minimal changes in device performance, indicating resilience to lower doses [41]. Further studies with gamma radiation doses up to 1 Mrad (Si) on both $\beta$-$Ga_2O_3$ SBDs and NiO/$\beta$-$Ga_2O_3$ heterojunction diodes indicated an intrinsically higher resistance to gamma irradiation [41-44]. While further investigation into the response of $\beta$-$Ga_2O_3$ to high doses of gamma radiation is of great interest due to its inherent radiation hardness, a systemic investigation into the impacts of such radiation on the electrical characteristics of β-Ga₂O₃ SBDs, particularly at higher cumulative doses, is still lacking.

In this study, we investigated the influence of gamma radiation on the electrical performance of β-Ga₂O₃ SBDs subjected to cumulative doses up to 5 Mrad (Si) using a $^{60}$Co gamma ray source. Through comprehensive electrical characterizations of the diodes both before and after different doses of gamma irradiation, we have gained valuable insights into β-Ga₂O₃'s resilience and potential degradation mechanisms under such extreme radiation conditions. Our findings highlight the exceptional tolerance of β-Ga₂O₃ SBDs to high gamma radiation doses, with minimal changes in electrical characteristics, including stable on-current, on-resistance, ideality factors and reverse breakdown characteristics even after exposure to high doses of 5 Mrad (Si).

## II. Experimental Section

The β-Ga₂O₃ Schottky barrier diode was fabricated on commercially available halide vapor phase epitaxy (HVPE) grown (001) n⁻ Ga₂O₃ film with ~10.8 μm thickness (Si-doped, $\sim 3 \times 10^{16}$ cm⁻³) on a ~625 μm thick Sn-doped ($\sim 5.4 \times 10^{18}$ cm⁻³) Ga₂O₃ substrate. A schematic cross section of the Schottky diode structure is shown in Figure 1(a). The device fabrication began with BCl₃-based reactive-ion etching (RIE) of the backside, where a ~1 μm thick layer of Ga₂O₃ was etched. This was followed by the deposition of a Ti/Au Ohmic metal stack using electron beam evaporation and rapid thermal annealing (RTA) in a nitrogen atmosphere at 470°C for 1 minute. Subsequently, the top Ni/Au Schottky contacts were patterned using electron beam lithography. Post-fabrication, current density-voltage (J-V) measurements were conducted with an HP 4155B semiconductor parameter analyzer. Additionally, a room temperature reverse breakdown measurement was performed, and reverse-biased C-V measurements on the Schottky contacts were carried out with an Agilent 4294A precision impedance analyzer. The gamma irradiation experiments were carried out at University of Massachusetts Lowell Radiation Laboratory using a $^{60}$Co γ-ray source with 1.173 MeV and 1.332 MeV energies at a rate of 1 Mrad (Si)/hour. The energy of the incident $^{60}$Co radiation ensures complete penetration of γ-rays into the material. The Schottky barrier diode was radiated for 1, 2, and 2 hours in successions at room temperature, which corresponded to cumulative absorbed doses of 1, 3, and 5 Mrad (Si), respectively. No bias was applied to the devices during irradiation to isolate and evaluate the effects of radiation exposure on device performance without interference from electrical stresses. The electrical characterizations (C-V and I-V) were performed after each cumulative dose of radiation exposure. The device responses were measured within 15 minutes following each irradiation dose to minimize any unintended annealing effects during the testing process, after which the samples were promptly returned to the irradiation chamber.

### III. Results and Discussions

The current-voltage (J-V) characteristics and specific on-resistance ($R_{on, sp}$) of the device under various cumulative doses of gamma radiation are shown in Figures 1(b)-(f). Initially, after exposure to 1 Mrad of irradiation, the on-current drastically decreases, while the on-resistance significantly increases. This degradation in on-current and increase in on-resistance are consistently observed across all investigated devices with different areas, as shown in Figure 1. This deterioration could be attributed to the generation of vacancy-related point defects, particularly Ga vacancies, [38, 39] which act as deep-level acceptor traps at the bulk and Schottky barrier interface during the initial ionizing radiation dose, thus reducing the electrical conductivity and forward current of the devices. A similar reduction in on-current was also observed in previous work on NiO/$\beta$-$Ga_2O_3$ heterojunction diodes after 1 Mrad of gamma radiation exposure [42]. However, upon exposure to higher doses of gamma radiation (3 and 5 Mrad), the on-current of the device increased, nearing pre-radiation levels, as shown in Figure 1 for different devices. Similarly, the specific on-resistance also returned to values close to the pre-radiation state. The recovery after 5 Mrad was substantial, with values ranging from 92% to 99% of pre-radiation performance, indicating that higher doses of gamma radiation successfully restored much of the device functionality degraded by the initial 1 Mrad exposure. This effect, known as 'radiation annealing,' occurs as gamma radiation generates secondary electrons through Compton scattering [33-35]. These electrons dissipate kinetic energy as heat throughout the lattice, annealing the trapping centers responsible for performance degradation. As a result, high cumulative doses of gamma radiation with high dose rates over prolonged exposure time leads to the restoration of device performance after initial degradation. Based on our current findings from this study, the radiation annealing effect occurs after total cumulative gamma doses of 3 and 5 Mrad (3 and 5 hours of exposure, respectively) at a relatively higher dose rate of 1 Mrad/hr (~277.8 rad/s). While previous reports on $\beta$-$Ga_2O_3$ SBDs have also indicated that annealing can occur at lower dose rates with prolonged exposure time, such as 1 Mrad radiation accumulated over 5 hours at 50 rad/s [44], a cumulative gamma dose of 1 Mrad over 6 hours did not induce radiation annealing in NiO/$\beta$-$Ga_2O_3$ heterojunction diodes [42], suggesting that device structure may also play a significant role in determining whether radiation annealing occurs. Additionally, studies on GaAs-based devices suggest that gamma radiation effects can also be influenced by pre-existing structural imperfections [34], indicating that factors such as total cumulative doses, incident gamma photon energy, dose rates, exposure time, as well as device structures all may influence the radiation annealing effect. For instance, while a recent study on NiO/$\beta$-$Ga_2O_3$ heterojunction diodes showed restoration of device performance after applying short forward current pulses during repeated I-V measurements at room temperature by annihilating radiation-induced trapped charges through carrier injection [42], our investigation into $\beta$-$Ga_2O_3$ SBDs indicates that device recovery can also occur after high dose of gamma exposure at higher dose rates (~277.8 rad/s) by the radiation annealing effect. Similar recovery of Schottky interface trap states due to radiation annealing has been observed in $\beta$-$Ga_2O_3$ and GaAs Schottky barrier diodes [33, 44].

To investigate the effects of different doses of gamma radiations on device capacitance, built-in-potential and net carrier concentration, C-V characteristics were performed after various cumulative radiation doses as shown in Figure 2. A reduction in capacitance is observed after exposure to 1 Mrad of radiation as shown in Figure 2(a) and Figure S1 of the supplementary material. However, after exposure with higher doses (3 and 5 Mrad), the capacitance was found to increase as the C-V curve shifts toward the pre-radiation condition. This behavior mirrors the I-V characteristics in Figure 1, where initial performance degradation is followed by a return to pre-radiation performance at higher doses. Similarly, after initial 1 Mrad of irradiation, the built-in-voltage increases from 0.95 V to 1.02 V, but subsequent higher doses of 5 Mrad restore the built-in potential to approximately 0.95 V, as depicted in the $1/C^2$-V plot in Figure 2(b). The built-in-voltage and net carrier concentration ($N_d^+ - N_a^-$) are determined using the following formulas, with the dielectric constant, $\varepsilon_0 = 8.85 \times 10^{-12}$ F/m in vacuum, relative permittivity of $\beta$-$Ga_2O_3$, $\varepsilon_r = 10$, density of states in the conduction band, $N_C = 5.2 \times 10^{18}$ $cm^{-3}$, where A representing the area of the device [45-47].

$$N_d^+ - N_a^- = \frac{2}{q\varepsilon_r\varepsilon_0 A^2 \left(\frac{d\frac{1}{C^2}}{dV}\right)} \qquad (1)$$

$$\frac{A^2}{C^2} = qV_{bi} + \frac{kT}{q} In\left[\frac{N_c}{N_d^+ - N_a^-}\right] \qquad (2)$$

The carrier concentration versus depth profile as shown in Figure 2(c), obtained from the C-V curves, reveals a small reduction in net carrier concentration, decreasing from ~$3.2 \times 10^{16}$ cm$^{-3}$ before radiation to ~$3.05 \times 10^{16}$ cm$^{-3}$ following irradiation. This reduction in carrier concentration indicates the formation of radiation induced acceptor-like deep level traps due to the radiation exposure. These traps effectively compensate free carriers available for conduction, leading to an overall reduction in net carrier concentration. This observation aligns with previously reported results for β-Ga$_2$O$_3$ SBDs as well as β-Ga$_2$O$_3$/NiO heterojunction diodes under neutron and gamma radiations [20-22, 42], where similar reductions in carrier concentration have been attributed to the introduction of radiation-induced deep level defect states. The carrier removal rates, which quantify such reduction, can be calculated using the following formula [42]:

$$R_c = \frac{n_{s0} - n_s}{\emptyset} \qquad (3)$$

Where, $n_{s0}$ is the initial carrier concentration, $n_s$ is the carrier concentration after radiation exposure and $\emptyset$ is the effective gamma radiation fluence. The effective gamma-ray fluence can be derived from the total ionizing dose using the conversion factor of 1 rad (Si) = $2.0 \times 10^9$ photons/cm$^2$ [41]. Small carrier removal rates of 0.605, 0.211, and 0.136 cm$^{-1}$ were observed after 1, 3, and 5 Mrad of cumulative doses, respectively, as summarized in Table 1. This low removal rate is due to the fact that gamma rays, while generating secondary electrons, can primarily induce displacement damage through non-ionizing energy loss [27]. Compton scattering of the primary $^{60}$Co gamma photons, with energies of 1.173 MeV and 1.332 MeV, may produce a high density of lower-energy photons ($E_\gamma \leq 0.60$ MeV). These Compton electrons lead to the creation of Frenkel pairs and defect clusters. Many of these defects can migrate, recombine, or form stable complexes that persist even at room temperature, thereby contributing to the observed low carrier removal rates [34]. Notably, the carrier removal rates of the irradiated SBDs show a general declining trend as the radiation doses increase as listed in Table 1, consistent with trends previously observed in neutron and gamma radiation studies [21, 41]. The carrier removal rates observed in this investigation after radiation exposure are also consistent with previously reported rates of $0.5 \pm 0.2$ cm$^{-1}$ and $0.007 \pm 0.001$ cm$^{-1}$ for absorbed doses of 1 or 100 kGy (Si), respectively [41]. In general, carrier removal rates due to gamma radiation are several orders of magnitude lower than those caused by protons and are 5-10 times lower than those caused by neutrons and electrons [21, 23, 41].

Temperature-dependent I-V-T measurements were also performed after 5 Mrad irradiation over a temperature range of 25˚C to 250˚C to investigate the effects of radiation at elevated temperatures as shown in Figure 3. In the forward bias region of the log-scale I-V graph in Figure 3(a), an increase in on-current with rising temperature is noted in the barrier-limited region (lower current density), which is expected from thermionic emission current transport. Conversely, in the ohmic region (higher current density), as shown in Figure 3(b), the current density decreases with increasing temperature due to reduced electron mobilities in the current-limiting series resistor at higher temperatures [48, 49]. This trend of increasing on-current in the lower current density region and decreasing current density in the higher current density region persists even after 5 Mrad radiation exposure, although the current density after radiation for all investigated temperatures is found to be lower compared to the pre-radiation conditions. Additionally, a gradual increase in leakage current is seen with higher temperatures, as shown in Figure 3(a), whereas after gamma radiation exposure, leakage current decreases at all investigated temperatures, with the reduction being more pronounced at elevated temperatures. The increase in leakage current at higher temperatures can be attributed to the electrons gaining higher energies, allowing them to more easily overcome the metal-semiconductor Schottky barrier, resulting in higher leakage current. This behavior is commonly observed in $\beta$-$Ga_2O_3$ SBDs and has also been noted in previous experiments [46, 48-50]. The specific on-resistance, $R_{on,sp}$ extracted from the forward I-V curve for different temperature indicates a steady increase with rising temperature as shown in Figure 3(c). This trend continues even after 5 Mrad of radiation exposure, where $R_{on,sp}$ consistently rises. The $R_{on,sp}$ as a function of temperature as shown in Figure 3(d) shows the steady rise in on-resistance, with the slope of this increase at elevated temperatures becomes steeper after radiation exposure. The higher on-resistance at elevated temperatures can be attributed to increased lattice vibrations, which decrease electron mobility and simultaneously reduce both the forward and reverse currents [21, 46].

The rectification ratio ($I_{on}/I_{off}$) of the diode, which reflects the behavior of forward and reverse currents at different temperatures determined at $\pm 4$ V bias voltage is plotted in Figure 4(a). The plot shows a steady reduction in the rectification ratio as temperature increases. While the ratio remains almost similar at room temperatures for both pre- and post-radiation conditions, it tends to increase at elevated temperatures after radiation, as the reduction of leakage current was more prominent for higher temperature. The ideality factor ($\eta$) and Schottky barrier height (SBH) extracted by considering the standard thermionic emission (TE) model [21, 46, 47] from the measured temperature dependent I-V curves (Fig. 3(a)) are shown in Figure 4(b).

$$J = J_s \left[ \exp \left( \frac{qV}{\eta k_0 T} \right) - 1 \right] \qquad (4)$$

$$J_s = A^* T^2 \exp\left(-\frac{q\Phi_B}{k_0 T}\right) \tag{5}$$

$$A^* = \frac{4\pi q m_n^* k_0^2}{h^3} \tag{6}$$

where q is the electric charge, $k_0$ is the Boltzmann constant, and η is the ideality factor, $J_s$ is the reverse saturation current density, $\Phi_B$ is the Schottky barrier height, and A* is Richardson's constant, which is calculated to be 41.04 A cm$^{-2}$ K$^{-2}$ [46, 49, 51]. Figure 4(b) shows a slight increase in the ideality factor after irradiation across all investigated temperatures, although temperature itself does not significantly impact the ideality factor. This slight increase of the ideality factor after irradiation can be attributed to trap defects generated by high radiation exposure. Additionally, the Schottky barrier height is observed to be highly temperature-dependent, gradually increasing with rising temperature, as shown in Figure 4(b). After irradiation, the SBH increases more noticeably at elevated temperatures, whereas the change between pre- and post-radiation conditions is minimal at lower temperatures. This temperature dependence of the SBH can arise from the lateral inhomogeneity of the barrier height [50]. Lateral inhomogeneity typically arises during the metal contact deposition process, which can introduce damage and disorder at the metal-semiconductor interface. As a result, the atomic structure and Schottky barrier deviate from the ideal homogeneous and abrupt interface [52]. Achieving a perfect epitaxial metal-Schottky interface is only feasible when the metal and semiconductor crystals exhibit near-perfect alignment in their two-dimensional lattice structures. As discussed in previous studies [50, 53], such lateral inhomogeneities may result in a temperature-dependent behavior of the SBH. According to the thermionic emission model, the ideal MS interface should be atomically flat and spatially homogeneous. However, in practice, a non-ideal inhomogeneous Schottky barrier often consists of locally non-uniform regions having lower and higher barrier height patches at the nanoscale. Due to these barrier inhomogeneities, the current conduction at the interface is not same at the whole temperature range [54, 55]. At lower temperatures, current conduction is due to carriers which cross the patches having relatively lower barrier heights, while at higher temperatures current conduction is dominated by those carriers which cross the patches having relatively higher barrier heights, effectively raising the overall SBH at elevated temperature. Such temperature-dependent increase in SBH has also been observed in other semiconductor-metal Schottky contacts [50, 54] and $Ga_2O_3$-based Schottky diodes [46, 56, 57]. The observed increase in Schottky barrier height following high doses of gamma radiation exposure might be attributed to the formation of acceptor-like compensating defects at the metal-semiconductor interface, leading to a reduction in carrier density and consequently elevate the barrier height. Such small increases in Schottky barrier height have also been observed in previous studies with neutron radiation [20]. With minimal changes in barrier height and ideality factor, particularly at lower temperatures, the small decrease in reverse leakage after radiation as shown in Figure 3(a) can be attributed to the lowering of the surface electric field due to the reduction in net carrier concentration from radiation damage [20, 58]. This reduction likely decreases the reverse leakage component from thermionic field emission, as expected by theoretical models [59]. However, at higher temperatures, the barrier height increases after radiation as shown in Figure 4(b), leading to a relatively higher reduction in leakage current (Figure 3(a)). While this reduction in leakage current is similar to behaviors observed in proton and neutron irradiated devices [20, 22, 25], further investigation is needed to fully understand these changes.

Finally, to assess the impact of high-dose gamma radiation on the reverse breakdown characteristics of the $\beta$-$Ga_2O_3$ SBDs, breakdown measurements were conducted before and after exposure to 5 Mrad of radiation. Figure 5 shows the breakdown characteristics of the diodes for multiple devices with same sizes for better comparison. It is observed that the reverse breakdown voltage ($V_{br}$) slightly increases from around 260 - 280 V in pre-radiated condition to 300 - 320 V under the influence of gamma radiation for all measured devices. A similar trend of increased $V_{br}$ after radiation has also been reported for both neutron and proton irradiated $\beta$-$Ga_2O_3$ SBDs [20, 22, 25] as well as for $SiO_2$ passivated 4H-SiC SBDs following 4 Mrad gamma radiation, which was attributed to an increase in negative interface charge [30]. While the increase in breakdown voltage after radiation can also be potentially related to the reduction of net carrier concentration due to the generation of compensating acceptor like bulk traps, further investigations are required to fully understand such behavior.

## IV. Conclusion

In summary, we conducted a systematic investigation of the electrical characteristics of $\beta$-$Ga_2O_3$ SBDs subjected to high doses of $^{60}Co$ gamma radiation, up to 5 Mrad (Si). Our findings reveal that $\beta$-$Ga_2O_3$ SBDs exhibit notable resilience to gamma radiation, with performance showing a complex interplay between degradation and recovery mechanisms. Initially, exposure to 1 Mrad gamma radiation leads to a significant reduction in on-current and an increase in on-resistance due to the formation of deep-level acceptor traps within the material. However, upon exposure to higher doses of 3 and 5 Mrad, the device performance demonstrated a partial recovery, likely due to a radiation annealing effect. A slight decrease in net carrier concentration following 5 Mrad (Si) irradiation is observed. Temperature-dependent I-V measurements showed a reduction in both forward and reverse currents post-irradiation, along with increased on-resistance, ideality factors, and Schottky barrier heights. Slight increase in the reverse breakdown voltage of the diodes is also observed after radiation, indicating impressive durability against gamma radiation, with inherent self-recovery processes mitigating the impact of high radiation doses. This resilience emphasizes the suitability of β-Ga$_2$O$_3$ based devices for applications in radiation-hardened electronics, particularly in extreme environments subjected to high levels of ionizing radiation.

**Supplementary material**

See the supplementary material for device C-V characteristics before and after radiation with various cumulative doses and a summary table containing the SBD parameters extracted from temperature-dependent J-V characteristics at different temperatures.

**Acknowledgments**

We acknowledge the funding support from AFOSR under award FA9550-18-1-0479 (Program Manager: Ali Sayir), NSF awards ECCS 2019749, 2231026, 1919798, ARPA-E award DE-AR0001879 and Coherent II-VI Foundation Block Gift Program. We also acknowledge the support and resources of UMass Lowell Radiation Laboratory in conducting the radiation experiments.

**Data Availability**

The data that support the findings of this study are available from the corresponding author upon reasonable request.

**Figure Captions**

**Figure 1.** (a) Schematic cross-section of the β-$Ga_2O_3$ Schottky barrier diode (b-f) J-V characteristic and specific-on-resistance of the diodes with varying areas after cumulative gamma radiation doses of 1, 3, and 5 Mrad (Si).

**Figure 2.** (a) Capacitance-voltage (C-V) characteristics for different cumulative doses of gamma radiation, (b) $1/C^2$-V plot used to determine the built-in potential ($V_{bi}$), and (c) Net carrier concentration profile as a function of depth, derived from the C-V curves.

**Figure 3.** (a) Temperature dependent J-V characteristics before and after 5 Mrad gamma irradiation in (a) log and (b) linear scales, (c) Specific-on-resistance vs. voltage characteristics of the diodes for different temperature, (d) Comparison of the specific-on-resistance as a function of temperature for pre- and post- radiation conditions.

**Figure 4.** (a) Rectification Ratio ($I_{on}/I_{off}$) and reverse leakage current of the diode at bias voltage of ±4 V, measured before and after 5 Mrad radiation, as a function of temperature, (b) Ideality factor ($\eta$) and Schottky barrier height (SBH) of the diode at various temperatures, both before and after irradiation.

**Figure 5.** Reverse J-V characteristics of the devices showing breakdown voltages before and after 5 Mrad gamma irradiation.

**Table 1.**

Summary of β-Ga$_2$O$_3$ Schottky barrier diode parameters before and after different cumulative doses of gamma radiation.

| Device Parameters | Pre-radiation | 1 Mrad | 3 Mrad | 5 Mrad |
|---|---|---|---|---|
| **Rectification Ratio, $I_{on}/I_{off}$** | $2.19 \times 10^{10}$ | $4.85 \times 10^{9}$ | $2.22 \times 10^{10}$ | $2.55 \times 10^{10}$ |
| **Built in Voltage, $V_{bi}$ (V)** | 0.96 | 1.02 | 0.957 | 0.956 |
| **Net Carrier Concentration, $N_D^+$-$N_A^-$ (cm$^{-3}$)** | $3.194 \times 10^{16}$ | $3.073 \times 10^{16}$ | $3.067 \times 10^{16}$ | $3.058 \times 10^{16}$ |
| **Carrier Removal Rate, $R_c$ (cm$^{-1}$)** | -- | 0.605 | 0.211 | 0.136 |

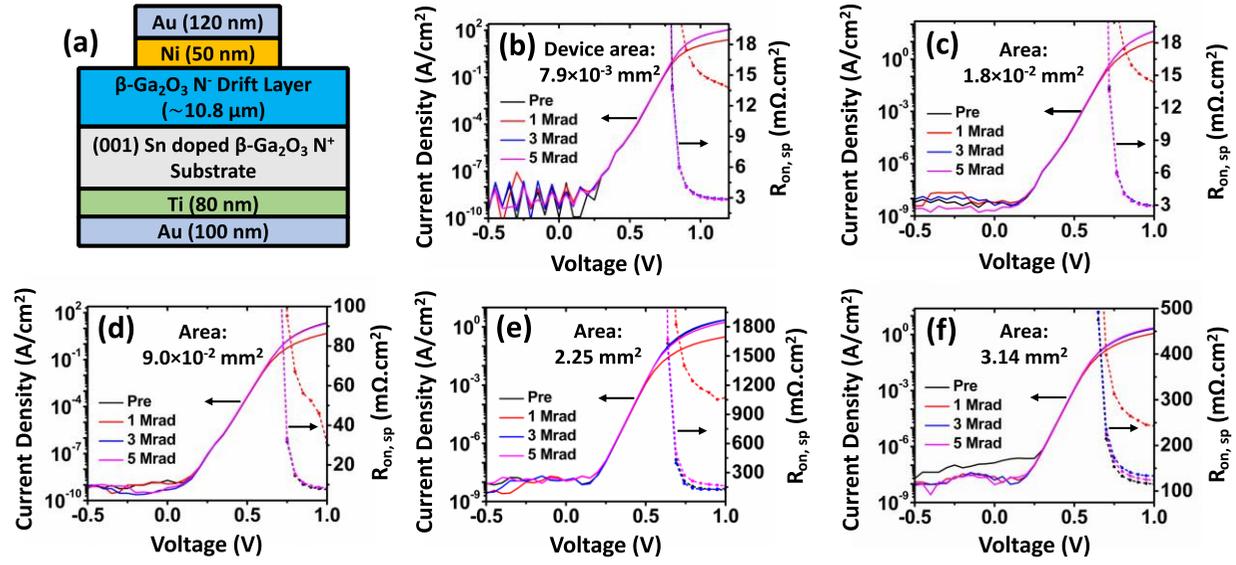

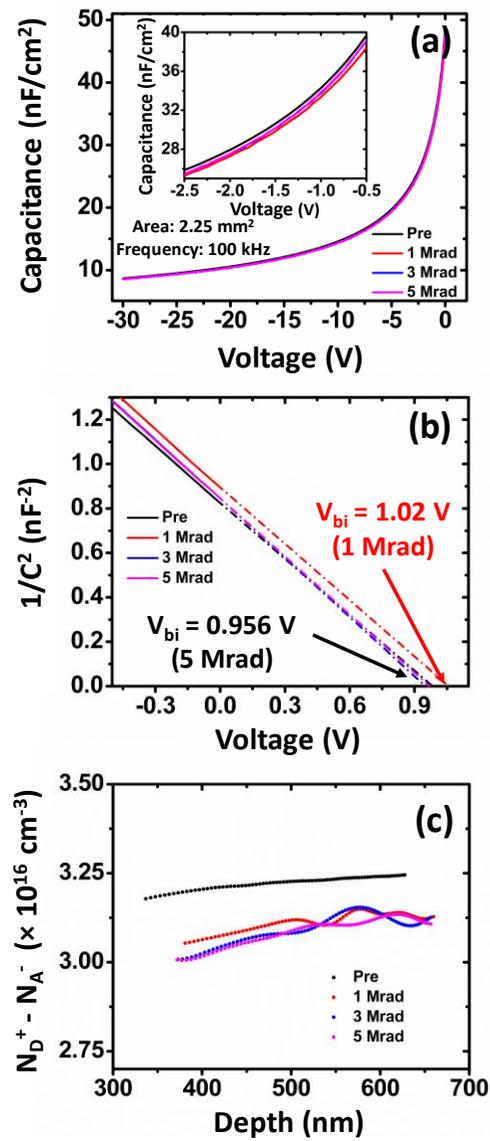

**Figure 3**

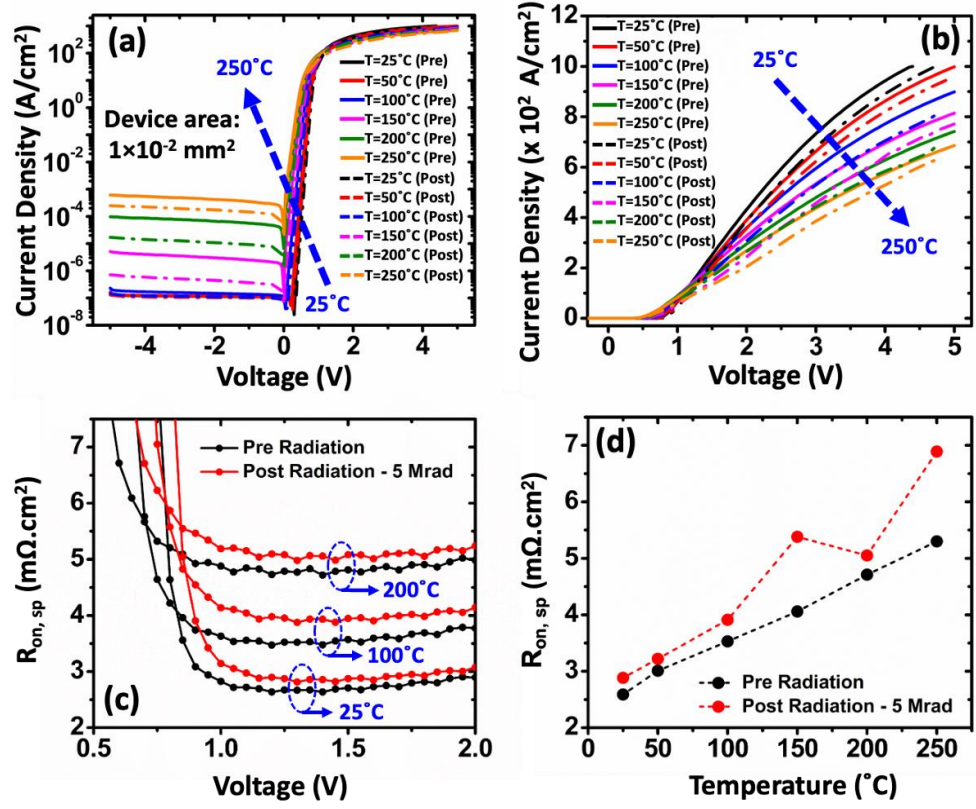

**Figure 4**

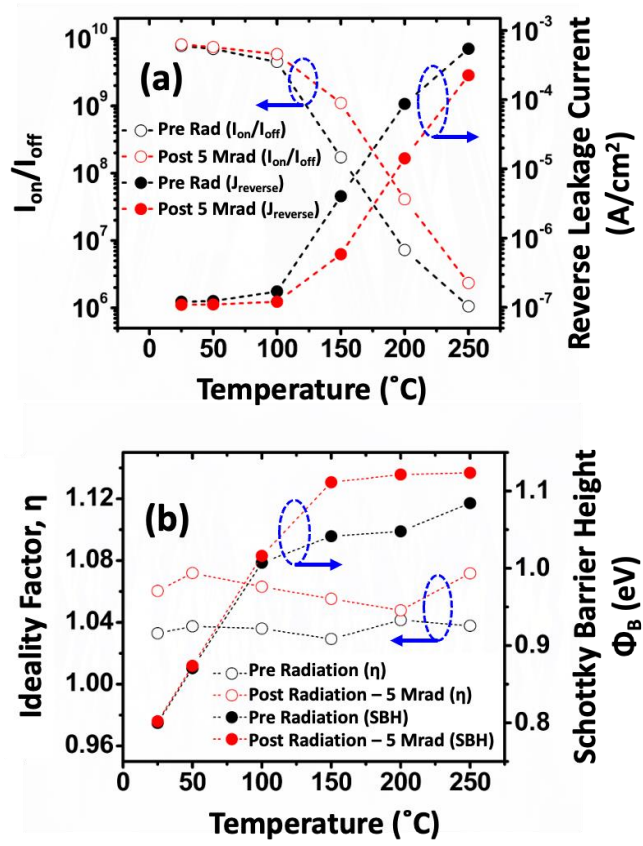

**Figure 5**

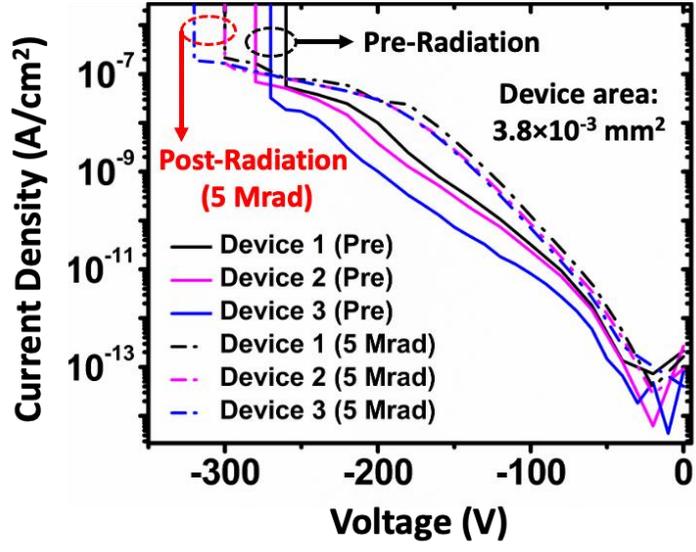